# Random Walk and Balancing[1]

*Frank Borg. Chydenius Institute, Jyväskylä University, 67101 Kokkola-Karleby, Finland*[2].

***Problem statement***: *What is the role, nature and cause of the apparent randomness exhibited by the neuromuscular/sensimotor system and observed especially in quiet standing and stick balancing?*

*Physical Review Letters* 1994 contained an intriguing paper titled "Random walking during quiet standing" (Collins and De Luca 1994). It dealt with the interesting phenomenon, that when you stand "still" (quiet standing) your body sways back and forth. Without external support it is impossible to stand like a rigid pillar. The most simple non-trivial physical model one can make of the situation is the inverted pendulum. The ankle joints are the pivot points and the muscles act like a spring keeping the pendulum upright. David A Winter *et alii* (1998, 2001, 20003) have shown how surprisingly well such a model accounts for some basic features of the quiet standing, if one excepts the "random walk" aspects of the stance[3]. In the present review we describe a few models of stance and balancing with special emphasis on trying to understand the role of the random behaviour. Studies of chaos in recent decades have teased us to look at "noise" as something probably more than just a "disturbance" (which one had hoped to be able to neglect while preserving the essential elements of the system). This shift of perspective has also accelerated due to insights from theories and experiments around stochastic resonance (SR) and self-organized criticality (SOC).

An electronic stabilometer consists of a rectangular (there are also triangular versions) plate with force transducers in the corners which measures the vertical forces. Using these data we can compute the center of pressure (COP) – see fig 1 which shows a top view of force plate – according to

$$x = \frac{a}{2} \cdot \frac{F_2 + F_3 - F_1 - F_4}{F_2 + F_3 + F_1 + F_4}$$
$$y = \frac{b}{2} \cdot \frac{F_1 + F_2 - F_3 - F_4}{F_2 + F_3 + F_1 + F_4}$$

where $F_i$ is the force measured by transducer #*i*. According to the conventions the person is standing such that the *y*-axis will be the forward direction (anterior/posterior = A/P) while the *x*-axis is along the sideway direction (medial/lateral = M/L). Nowadays the force plates are of course interfaced with computers. The measurements procedures are simple but require standardized test conditions in order to ensure repeatability. In our static tests we have adopted the "Morton"-foot position with feet positioned with a 30 degree angle between medial sides and a 2 cm heel-to-heel separation. The arms are relaxed on the sides and the gaze fixed on a point 3 m away at 1.5 m height. Typical measurement time used is 60 seconds. Disturbing sounds, movements and lights must naturally be avoided. For standardization issues see Kapteyn *et al.* (1983), Browne & O'Hare (2000), Carpenter *et al.* (2001) and McIlroy & Maki (1997).

---

1  This is an English version of the paper "Stillstående slumpvandring", *Arkhimedes* (Helsinki-Helsingfors) 4, 19-28 (2003) (in Swedish).
2  E-mail: borgbros@netti.fi .
3  The passive stiffness control model proposed by Winter *et alii* has however been criticized by a number of investigators. These aspects will be dealt with in a subsequent paper which draws e.g. on EMG-measurements which demonstrate how the *gastrocnemius* muscles modulate the balancing.

The force values are transformed to COP-coordinates which produces two time series, $x(t)$ and $y(t)$, so called *stabilograms*. If we draw the curve $(x(t), y(t))$ in the plane we obtain the so called *statokinesigram* (or *posturogram*) which shows how the COP shifts with time (fig. 2). A central question that has been addressed during the last decades is how to extract information from these time series that might have diagnostic relevance. A related question is naturally whether the data can be successfully interpreted in terms of a mathematically formulated neuro-muscular model.

These questions hint at two main types of approaches: on the one hand we have the classical Newtonian methods – physical modelling – which is based on the physically motivated mathematical models of the phenomena; on the other hand we have the so called algorithmic modelling (Rapp *et al.* 1999, Jones 2001). In the algorithmic modelling approach one does not try to model the mechanism in itself which is instead considered as a black or gray box; it is the data (the time series) which is the main object whose characteristics is to be explained using a mathematical modelling. System identification is a discipline that can be said to have much in common with algorithmic modelling. An interesting link between these two main approaches is provided by recent works on non-linear time series analysis which originated with the study of non-linear dynamics and chaotic systems (Abarbanel 1995, Kantz and Schreiber 2000). These studies have revealed that under quite general circumstances it is possible to extract a number of characteristic invariants of the dynamical systems without a detailed knowledge of the system, and only using the data of the time series generated by the system. One example of such an invariant is the topological dimension of the system or its "attractor". Another point is that these non-linear methods are not sensitive to how the data is represented because one is looking for invariants. Thus, global Ljapunov-exponents are independent of the coordinate system used. Other interesting parameters introduced via non-linear time series analysis are the various entropies (Kolomogorov-Sinai, Renyi, Approximate Entropy), correlation dimension, mutual information, etc. Another observation is that for complicated system (such as living organisms) it happens that a large number of the degrees of freedom get "frozen" and that the system in certain aspects thus behaves as a low-dimensional dynamical system. Hermann Haken and his associates have discussed a number of instances of such a "freezing" of the degrees of freedom (Haken 1983), a phenomenon that gives reason to think that non-linear time series analysis may find uses in biomedical signal analysis[4].

A quite common idea in chaos research is that "healthy" organic systems may have a number of chaotic properties; indeed, a chaotic dynamics implies that the system probes an extended set of points in the phase space and thus is ready to react to new unexpected circumstances, while a too regular system remains locked up in small part of the phase space. There are experimental indications of a higher correlation between A/P- and M/L-swaying for MS and Parkinson patients than for normal participants (Rosenblum *et al.* 1998). One may assume that a more rigid coupling between the A/P- and M/L-swayings is associated with a reduced ability to correct posture when subjected to perturbations.

---

4   In quiet standing we have an obvious example of the "freezing of degrees of freedom" in the sense that by the co-activation of a number of postural muscles the standing person can to a good approximation be modelled as a human inverted pendulum.

The system of postural control involves the vestibular system which registers the linear accelerations and rotations of the head; skin receptors; the eyes; joint receptors; muscle spindles which register length and the rate of contraction of the muscle; Golgi-organs at the muscle tendons which gauge the tension. The feet are especially well endowed with ligaments and receptors. Standing on a soft compliant material affects the swayings significantly. Signals from the receptors in the muscles and ligaments are routed via synapses in the spine to motor neurons which control the muscles, thus creating a feedback system. The typical response time for leg muscles (gastrocnemius, soleus, peroneus longus, etc) is about 90 – 140 ms. The postural feedback mechanism can e.g. be tested by stimulating neck muscles which will prompt automatic postural reflexes.

The clinical study of stance is called posturology (Gagey and Weber 1999). Quite a number parameters have been used over the years in order to describe the posturograms. Chiari *et al.* (2002, 2000) list 55 parameters, many of which of course are overlapping. These authors conveniently classify the parameters into three groups:

- time-based parameters
- frequency-based parameters
- stochastic parameters

Baratto *et alii* (2002) present another classification dividing the parameters into two categories: *global parameters* and *structural parameters*. The global parameters are described as those "which estimate the overall 'size' of the sway patterns", whereas structural parameters are those "which attempt to decompose the sway patterns into elements and then examine their interactions". Trace-length and diffusion analysis parameters (discussed below) are given as examples of the respective categories. One of the conclusions of the study by Baratto et *alii*, after investigating a set of 39 parameters, is that the set can be basically reduced to two global and two structural parameters that are particularly useful in the clinical practice (more below).

The group of time-based parameters includes descriptive parameters such as the area covered by the COP-points (can be defined in various ways); the length of the COP-trace (depends of the resolution); average velocity in the A/P- and the M/L-direction. The power-spectrum of the stabilograms (range typically 0 – 10 Hz) yield frequency-based parameters such as the average frequency. The investigations Collins and De Luca (1993, 1994) brought the stochastic parameters into the picture. Within the field of posturology the stochastic study is often referred to as "stabilographic diffusion analys" (SDA) as there seems to be an analogue with the diffusion as a stochastic process. Collins and De Luca made an attempt to interpret the "spaghetti curve" of the posturogram as a result of deterministic chaos:

"A likely candidate for physiological chaos is the human postural control system, the output of which is highly irregular .... The identification of postural sway as an instance of chaos would suggest that there is a simple, dynamical mechanism at work in balance regulation and may make possible new therapeutic and preventative strategies for postural instability" (Collins and De Luca 1994: 764).

In order to test the hypothesis of deterministic chaos as an explanation of the random looking aspects of the swaying, the authors estimated the Ljapunov exponents for both the original data and a phase randomized substitute data (which has the same statistical properties as the original data while all the deterministic correlations are wiped out) and they got the same results, concluding that: "(...) the postural control system should not be modeled as a chaotic process and that it is better represented as a stochastic one" (*ibidem* p. 765). It might be cautioned that the usual estimations of the Ljapunov exponents are quite sensitive to noise, and a noise level of 2 – 3% may be enough to render an estimation inconclusive (Kantz and Schreiber 2000).

The favourite stochastic models that have been used in this connection are the *fractal Brownian motion* (fBm) and the *fractal Gaussian noise* (fGn). Mandelbrot and Wallis (1968) coined these terms but the models were studied already around 1940 by A N Kolmogorov. A basic idea underpinning these models is that of a *self similar process*; or, more exactly, an *H-self similar process* $x(t)$, which is defined by the requirement that the scaled version $x(at)$ is statistically equivalent with the process $x(t)$ multiplied by a factor $a^H$ ($H$ is the so called *Hurst*-exponent),

$$(1) \quad Prob(x(t)<x) = Prob\left(\frac{1}{a^H} x(at) < x\right)$$

which is commonly symbolized by

$$(2) \quad x(t) \stackrel{d}{=} \frac{1}{a^H} \cdot x(at)$$

In order to characterize the fBm-process $x(t)$ we assume that $x(0) = 0$ (with probability 1), then the variance of the fBm satisfies

$$(3) \quad \langle (x(t+\Delta) - x(t))^2 \rangle = \sigma^2 |\Delta|^{2H}$$

For $H = 0.5$ we obtain the ordinary Brownian motion where the variance is a linear function of time $\Delta$. The cases $H \neq 0.5$ are referred to as *anomalous diffusion* (see e.g. Metzler and Klafter (2000)). It can be argued mathematically (Flandrin 1989) that the average power spectrum for fBm will be

$$(4) \quad S_x(f) = \sigma^2 \, \Gamma(2H+1) \sin(\pi H) \cdot \frac{1}{(2\pi f)^{2H+1}}$$

that is, the power spectrum depends on the frequency as $f$ raised to $-2H – 1$. Thus, the bigger the *H*-value the smoother we expect the curve $x(t)$ to be, since the high frequency components are more damped for bigger $H$. In fact, the graph of a *H*-self similar curve defines a set in the plane with the fractal dimension $D = 2 – H$.

In order to check whether the data is compatible with an fBm-process there are two obvious avenues suggested by (3) and (4): First, the method of variance based on equ (3) where we calculate the the variance for a series of time differences $\Delta$ and try to fit the result to a power law; secondly, the method of power spectrum where we make an FFT-analysis of the data and try to fit the estimated power spectrum to a $1/f$-spectrum (4). Collins and De Luca (1994) for instance employ the variance method whereas Thurner *et alii* (2002) also use the power spectrum method in order to analyze posturogram data. Both methods give similar results. The time as well as the frequency range tend to be divided into two regimes characterized by *persistence* ($t < 1$ s, $f > 1$ Hz) and *anti-persistence* ($t > 1$ s, $f < 1$ Hz). Persistence means that the movement tends to continue in the same direction as before, whereas anti-persistence means that the movement is likely to reverse. Collins and De Luca (1994) obtained for the short-time Hurst-exponent $H_s = 0.83 \pm 0.04$, and for the long-time Hurst-exponent $H_l = 0.25 \pm 0.06$. For time long enough the Hurst exponent should approach 0 according to (3) since the motion (swaying) is bounded.

The physical reason for these two regimes is quite clear. In the short-time perspective the body is likely to continue to move in the same direction as the moment before due to its inertia. However, when the body deviates by more than 0.5 degrees from the "equilibrium position" we expect a correction to take place which will temporarily return the body to the upright equilibrium position[5]. This process suggests a mechanical model (Peterka 2000) in the form of an inverted pendulum, with a spring that returns it to the vertical position, and a stochastic "disturbance" that explains the erratic swaying.

$$(5) \quad I\ddot{\alpha} + K_D\dot{\alpha} + (K_P - mgL)\cdot\alpha + K_I\int_0^t \alpha(u)\,du = \frac{1000}{\tau_f}\cdot\int_0^t e^{\frac{(u-t)}{\tau_f}} n(u)\,du$$

Here $\alpha$ stands for the angular deviation from the vertical in the A/P-direction (sagittal plane), $I$ is the moment of inertia with respect to the ankle joint (pivot point), $K_P$ is the spring constant ("stiffness"), $m$ is the mass of the pendulum (body mass minus mass of the feet), and $L$ is finally the distance from the ankle joint to the center of mass (COM). On the right hand side of (5) we have a Gaussian noise factor $n(t)$ with the variance 1 and mean 0, whereas the exponential factor represents a low pass filter. The model also incorporates a friction term and an "integrator" in accordance with the classical PID-control models for plants (Khoo 2000).

For the model (5) we can calculate the spectrum for the COM-coordinate $y_c$, which is proportional to the angle $\alpha$ ($y_c = L\alpha$ for small angles $\alpha$), and for realistic parameter values we indeed obtain two distinct regions around the "threshold" frequency (using parameter values from Peterka 2000)

$$f_n = \frac{1}{2\pi}\sqrt{\frac{K_P - mgL}{I}} \approx 0.42\,\text{Hz}$$

This threshold behaviour is smoothed out for the COP-coordinate because in the frequency space it is (formally) related to the COM-coordinate by

$$|\hat{y}(\omega)|^2 = \left(1 + \frac{\omega^2}{\omega_c^2}\right)^2 \cdot |\hat{y}_c(\omega)|^2$$

with

$$\omega_c = 2\pi f_c = \sqrt{\frac{mgL}{I}}$$

---

[5] In quiet standing humans lean a bit forward so that for the average position the center of mass will be ca 5 cm ahead of the ankle joints.

The previous modell used "white noise" ($H = 0.5$) for the disturbances, and the $H$-self similarity is shown to be only approximately valid within some restricted ranges of time and frequency. Chow and Collins (1995) who studied the balance using a pinned polymer model driven by stochastic noise realized that the *fluctuation-dissipation theorem* (FDT) of statistical physics would be applicable in such a model. With the FDT it is possible to predict the characteristic responses to small perturbations using only perturbation-free data (its auto-correlation function). This seems indeed to have been verified experimentally for quiet standing (Lauk *et al.* 1998; Hsiao-Wecksler *et al.* 2003). One implication is that it might be possible to predict the response of persons to small perturbations without actually subjecting them to physical perturbations.

Apparently (5) describes a continuous version of the ARMA-models ("Auto-Regressive-Moving Average") which are quite popular in the biomedical field (Bruce 2001; Rangayyang 2002). If we assume that the coordinate $x_i$ at time $t_i$ is a linear function of past values plus a stochastic term we get the equation

(6)    $$x_i = \sum_{k=1}^{n} a_k x_{i-k} + \sum_{k=1}^{m} b_k w_{i-k}$$

Neglecting the stochastic term this model goes under the name of *linear prediction* (LP) (see e.g. Press *et al.* 2002 §13.6). The figure 5 shows a curve of simulated data (upper curve) according to (6) as compared with real data (bottom curve). In the simulation we have used the values $n = 10$, $m = 40$, and the filter parameters $a$ have been calculated from the measurement data using an adaptive LMS-algorithm (the classical adaptive Widrow-Hoff algorithm is discussed e.g. by Haykin 1999 and Hänsler 1997). The stochastic terms $w$ were finally generated according to a Gaussian distribution with the standard deviation 1 and mean 0, while the MA-parameters were set to $b_0 = .... = b_{39} = 0.038$. Equ (5) and (6) may illustrate a typical difference between a physical model and an algorithmic model. Physical models have the obvious advantage of connecting the model parameters with physical characteristics that in principle can be measured and compared with the model. The ARMA-model on the other hand may be easy to fit to the experimental data. From a diagnostic point of view it may be enough to obtain parameter sets that are able to categorize the data into "normal" and "pathologic". This is also a typical objective for *Artificial Neural Networks* (ANN) which belong to a group of analysis methods termed "soft computing" (Tettamanzi and Tomassini 2001). Algorithmic modelling may be justified when there is no obvious physical model at hand or it is too complicated to be the basis for practical calculations.

Rosenblum *et alii* (1998) present a completely deterministic model of the swaying. The starting point is again the inverted pendulum, in this case supplied with a non-linear feed back control

(7)    $\ddot{\alpha} + 2h\dot{\alpha} + \omega^2 \alpha + c_1 F(\alpha(t-\tau), \lambda_1) + c_2 F(\dot{\alpha}(t-\tau), \lambda_2) = 0$

The function $F$ is thought to mimic the response of the proprioceptive system to the angle α and its rate of change $d\alpha/dt$. Their simplest example is a piecewise linear function (Θ denotes the Heaviside step-function)

(8)    $F(x, x_0) = (x - x_0)\Theta(x - x_0) + (x + x_0)\Theta(-(x + x_0))$

In general 1- and 2-dimensional autonomous continuous dynamical system do not generate chaos (result of the Poincaré-Bendixon-theorem; see e.g. Alligood *et alii* (1996)). However, with the delayed feedback control the situation might change; a classical example is the "Ikeda-oscillator" (Ikeda and Matsumoto 1987)

(9)    $\dot{x}(t) + \alpha x(t) + \beta \sin x(t - \tau) = 0$

which generates a chaotic signal (e.g. for α = 1, β = 20, τ = 2). Another classical illustration of chaos, via period doubling, is given by the logistic map

$$x_{i+1} = \lambda x_i (1 - x_i)$$

is in fact related to (9) (if we ignore the velocity term in (9)) and gives a hint why chaos may be expected for dynamical systems with time delayed feedback.

Milton (2000) presents some interesting ideas on the role of stochastic signals in biology though e.g. the model in (Eurich and Milton 1996) for balance may be somewhat too schematic for throwing much light on the phenomenon. It is also suggested that the phenomenon of *stochastic resonance* (SR, see e.g. Moss 2000) may play a significant role in many biological processes. A recent study (Kitajo *et al.* 2003) indicates that subliminal visual stimuli may improve the sensimotor sensitivity. Soma *et al.* (2003) have presented experimental results according to which 1/$f$-nose is more effective than white noise in enhancing the sensitivity. Another interesting finding comes from the group of J J Collins showing that subliminal tactile stimuli can slightly improve the score in the balance tests (quiet stance). In one experiment (Gravelle *et al.* 2002) electrical stimulation at the knee was used, in another experiment small vibrating nylon rods under the feet were employed (Priplata *et al.* 2002). The amplitude of the stimuli was tuned to a level where the participant no longer was consciously aware of it. One interpretation of these results is that the subliminal stimuli improves the sensitivity of the feedback control mechanism in line with the theory of SR. The Collins group have already tested vibrating soles aimed at persons with impoverished balance (as is the case with many elders).

An important point thus is that "noise" need not necessarily be considered as a detrimental "disturbance" but may be precondition for the optimal functioning of the sensory organs. However, this aspect of "noise" seems not to have been explicitly integrated in any of the published models of balance. However, for stick-balancing Cabrera and Milton (2002) have suggested that *parametric noise* could be a significant element of the balancing mechanism. Thought stick-balancing is not the same thing as quite stance, it might be of interest understanding quiet stance since in both cases some sort of a feedback system is involved, and in both cases the apparent noisy behaviour is a prominent feature. In the experiments a 62 cm long stick was balanced on the finger tip and two LED-markers on both ends of the stick were applied so that its position an attitude could be recored with the help of three video cameras. For a homogenous stick of length *l* and mass *m* whose pivot point only moves horizontally along a line we obtain the Newtonian equation (of the inverted pendulum) on the form

$$(10) \quad \ddot{\alpha} - \frac{3}{2}\frac{g}{l}\sin\alpha + \frac{3}{2l}\ddot{\xi}_0 \cos\alpha = 0$$

where $\xi_0(t)$ denotes the position of the pivot point (finger tip) along the horizontal *x*-axis, and $\alpha$ is the angle (attitude) of the stick in relation to the vertical direction (in the sagittal plane). In this model the motion of the stick is controlled by shifting the position $\xi_0(t)$, which is thus the "control variable". Cabrera and Milton suggests as a first approximation to set the control term proportional to the deviation $\alpha$ from the vertical direction. One important qualification is that the control term involves a time delay $\tau$ due to the sensimotor system. Thus, with rescaling, and an additional friction term, the equation for this model may be written as

$$(11) \quad \begin{aligned} &\ddot{\phi}(u) + \tau\Gamma\dot{\phi}(u) - \tau^2\omega^2\sin\phi(u) + R(u)\phi(u-1) = 0 \\ &\alpha(t) = \phi\left(\frac{t}{\tau}\right) = \phi(u) \end{aligned}$$

Here the time has been rescaled ($t \to u = t/\tau$) so that the delay $\tau$ appears as a factor of the frictional constant $\Gamma$ and the "eigen-frequency" $\omega/2\pi$ of the stick. *R* in the last rhs term sizes the magnitude of the restoring force. The factor $\tau$ does not appear here since this control term corresponds to the force and thus the acceleration of the hand (second order time derivative of $\xi_0(t)$). The central idea of the authors is the suggestion that the "force" *R* may be described, in accordance with the idea of "parametric resonance", as a sum of a constant part and a fluctuating part,

$$(12) \quad R(u) = R_0 + \eta(u)$$

where $\eta$ is supposed to represent Gaussian noise. (A somewhat related notion is that of "parametric oscillators", see e.g. Landau & Lifshitz 1976, §27; José & Saletan 1998, §7.3.) The implication is that muscular force is supposed to be fluctuating. Indeed, there are indications of spontaneous vibrations of the muscles (Basmaijan and De Luca 1985, p. 152). In the synapses, between motor neurons and muscles, there is a degree of chance operating in the secretion of transmitter substances such as acethylcoline.

Here we have to point out one problem with the paper by Cabrera and Milton (2002). They interpret the angle ($\alpha$ /$\phi$) in (10) and (11) as the angle ("latitude") which the stick makes with the vertical in 3D-space, not as the angle in the sagittal plane which is presupposed by the 2D physical model in (10-11) (where $\alpha$ is the angle of rotation around the y-axis if the x-axis is taken to be along the anterior/posterior direction and z-axis to be along the vertical direction). Whereas the latitude varies from 0 to 180 degrees, the sagittal angle $\alpha$ implied in (10) and (11) varies from -180 to 180 degrees, and average position of the stick may be supposed to be $\alpha = 0$. Since Cabrera and Milton have calculated the latitude angle from the data they naturally get a positive average angle. Indeed, if we assume that $\omega^2 \tau^2 > R_0$ then there is a non-zero stationary solution to (11) given by

$$\frac{\sin(\phi)}{\phi} = \frac{R_0}{\omega^2 \tau^2}$$

However, this solution is non-physical because it would imply a constant acceleration $\ddot{\xi}_0$ of the hand (see (10)). Thus, there is a mix-up between the interpretation of the angle and the physical model in their paper. As a contrast Mehta and Schaal (2002) which also analyze stick-balancing, clearly distinguish between the motion in the A/P- and in the M/L-direction. For small deviations from the vertical direction, equ (10) is a good approximation for the dynamics of the A/P-oscillations (the full 3D equations are a bit more complicated but reduce to (10) for the A/P-oscillations if the deviations from the vertical direction may be supposed to be small).

Still it might be of some interest to scrutinize (11) from the mathematical point of view. By varying $\tau$ and $R_0$ we can determine the region of stability in the ($\tau$, $R_0$)-plane. For small $\tau$-values a Taylor-development can be used with suggests that in this case (also neglecting the contribution by the noise) the region of stability is given by

(12) $\quad \tau \Gamma > R_0$
$\quad\quad R_0 > \omega^2 \tau^2$

More exactly, if we make the solution ansatz $e^{\lambda t}$ in the linearized version of (11) (replacing sin($\phi$) with $\phi$) and again neglect the noise, we obtain the characteristic equation

$$\lambda^2 + \tau \Gamma \lambda - \tau^2 \omega^2 + R_0 e^{-\lambda} = 0$$

whose zero-points in $\lambda$ with negative real parts correspond to the stable states (for an analysis in the case $\Gamma = 0$ see Atay (1999)). The interesting physical idea proposed by Cabrera and Milton (even if the model does not directly apply to stick-balancing in this form) is that the system tends to stay close to boundary of the region of stability. They propose that such a mechanism could shorten the effective reaction times of the feedback control and also explain the peculiar statistical properties they found in their data associated with intermittence (more on that below). We may add, that on physical grounds a quite obvious general strategy in controlling the stick could be to apply jerky motions; that is, short rapid accelerations/decelerations of the hand[6], since it has a quite a limited range of movement.

In a subsequent paper Cabrera and Milton (2002b) associated the tendency to stay close to the boundary of stability with the concept of self-organized criticality (SOC) advanced by Per Bak, Chao Tang and Kurt Wiesenfeld (1987) – for a review see Jensen (1998). The paradigmatic example is that of a heap of sand. Grain is added to grain, eventually small instabilities appear in the heap causing small avalanches that may now and then discharge larger avalanches. Analogously it has been proposed that many biological systems operate close to a critical instability thus enhancing their sensitivity. (For a discussion of SOC in relation to the brain see e.g. Linkenkaer-Hansen 2002.) In the case with the stick-balancing one might speculate that nearness to the in/stability region quickens the reactions, as suggested by Cabrera and Milton. They also propose (Cabrera and Milton 2002b) on the basis of their data that the time series of the latitude-angle of the stick exhibits the characteristics of a so called Lévy-flight (the concept discussed e.g. by Jespersen *et al.* 1999) which generalizes Brownian motion. A Lévy-process with the Lévy-parameter $0 < \mu < 2$ has a distribution which in the force-free case can be described by a differential equation (a special case of the fractal Fokker-Planck equations) of the form

$$(13) \quad \frac{\partial f(x,t)}{\partial t} = D_\mu \frac{\partial^\mu f(x,t)}{\partial |x|^\mu}$$

The fractal derivative in (13) may be defined in terms of Fourier-transformations (Riesz), or via fractal integration (see e.g. Saichev and Woyczyñski 1997 § 6.9). The limiting case $\mu = 2$ coincides with the ordinary Brownian motion which indeed has a Gaussian distribution. Equ (13) can be solved using Fourier-transformations. Given the boundary condition

$$f(x,0) = \delta(x)$$

the distribution will asymptotically approach (can also be guessed via scaling arguments)

$$(14) \quad f(x,t) \approx \frac{D_\mu t}{x^{1+\mu}}$$

According to Cabrera and Milton (2002b) the distribution of the change $\Delta v$ in the velocity, for a fixed time interval $\Delta t$, shows the characteristics of the Lévy-statistics. The distribution seems to be clearly different from Gaussian distribution and there is a significant difference also between beginners and experienced balancers. Beginners have a more truncated distribution; that is, a smaller proportion of big velocity changes. Furthermore it was found that the return probability density – the probability of the time interval between successive crossings of threshold angle – adheres to a power law

---

6  One could make a comparison with the so called "bang-bang" control in control theory.

(15)     $P(\delta t) \propto \delta t^{-\frac{3}{2}}$

The same distribution is found for ordinary Brownian motion too; indeed, for fractal time series we have in general (Ding and Yang 1995)

(16)     $P(\delta t) \propto \delta t^{-D} = \frac{1}{\delta t^{2-H}}$

where $D = 2 - H$ is the fractal dimension of the curve ($H = 0.5$ for ordinary Brownian motion). Heagy *et alii* (1994) have however also demonstrated that the exponent -3/2 is a universal characteristics of intermittency; that is, fractal processes with laminar phases interspersed with big excursions. Cabrera and Milton advances the idea that (15) implies that parametric noise enables swifter reactions than the typical neuromuscular delay around 100 ms seems to allow. Atay (2002) describes a class of simple delayed feedback control systems where asymptotic stability is possible only if the delay satisfies (homogeneous inverted pendulum of length *l*)

(17)     $\tau < \sqrt{\frac{2l}{3g}}$

For $\tau$ = ca 100 ms we get for the corresponding length 15 cm. Indeed, for most people it is probably impossible to balance an ordinary pen on the fingertip. However, if intermittency helps, could this be demonstrated by being able to balance short sticks than expected on the basis of reaction time alon?

The idea connected with model (11-12) is that the system is to live close to the edge of instability region whereby fluctuations may throw it into the instability region, initiating an excursion till another fluctuation maybe will bring it back to the stability region. Even if the model as such applied to the balancing is wanting, the discovery of intermittency and the Lévy-distribution in the stick balancing data warrants further study. Other similar experiments could be suggested, such as the balancing on a mono cycle (for a mathematical analysis see Johnson 2002). Mention should also be made of the more traditional theories of control, such as has been forwarded by Mehta and Schaal (2002) in connection with stick balancing. One idea being that the balancing mimics a Kalman-filter based control. The biological control system is assumed to predict the evolution and to use this prediction in order to compensate for the time delay in the control loop (Wolpert *et al.* 1995; Wolpert 1997). A classical "dead time" compensating control model is the so called Smith predictor (Smith 1957) and it has been proposed that the cerebellum harbors internal models (Dingwell *et al.* 2001) of movement patterns of the Smith type (Miall *et al.* 1993). An interesting recent study (Moreau and Sontag 2003) describes a general non-linear adaptive control mechanism, and it is suggested that it could explain the automatic fine tuning of the parameters in models of the saccadic movements of the eye and the oscillations of the cilia in the ear.

From the dynamical point of view an interesting query is whether quiet stance and stick balancing in some sense belong to the same "universality class". Anyway, the "noise" seems to be an integrated part of the neuromuscular system. In quiet stance a stream of signals is emanating from muscles, joints, skin, tendons, eyes, ears, etc. Sometimes the sensor inputs give contradictory information about balance and the body position. Also there is the phenomenon of adaption; a constant stimuli wears off. Perhaps the erratic swaying, seen from this perspective, may act as an updating mechanism of the "body-memory", analogous to the brain oscillations around 10 – 20 Hz that are hypothesized to keep the brain agile (Linkenkaer-Hansen 2002). The subliminal stimuli may enhance this updating mechanism, whereas too much noise masks the relevant information and perturbs the balance adversely.

*Acknowledgments*

This paper was written during a Biosignals-project whose main sponsor was the Finnish technology agency *TEKES*. Project manager was Mika Herrala, and the department head prof. Ismo Hakala, who together with Mats Manderbacka (CEO, Hur Co) were instrumental for the initiation of the project. I am also indebted to J G Milton for stimulating correspondence.

*Literature*

1   Abarbanel H D I: *Analysis of Observed Chaotic Data*. Springer 1995.

2   Alligood K T, Sauer T D and Yorke J A: *Chaos. An Introduction to Dynamical Systems*. Springer 1996.

3   Atay F M: "Balancing the inverted pendulum using position feedback". *Appl. Math. Lett.* 12 No. 5 (1999) 51 - 56.

4   Bak P, Tang C and Wiesenfeld K: "Self-organized criticality: An explanation of $1/f$-noise". *Phys. Rev. Lett.* 59 (1987) 381 – 384.

5   Baratto L et al.: "A new look at posturographic analysis in the clinical context: Sway-density vs . other parametrization techniques". *Motor Control* 6 (2002) 246-270.

6   Basmajian J V, De Luca C J: *Muscles Alive. Their Functions Revealed by Electromyography*. 5. ed. Williams & Wilkins 1985.

7   Browne J and O'Hare N: "Recette de plates-forme de force". *Physiol. Meas.* 21 (2000) 515 – 524.

8   Bruce E N: *Biomedical Signal Processing and Signal Modeling*. Wiley 2001.

9   Cabrera J L and Milton J G: "On-off intermittency in human balancing task". *Phys. Rev. Lett.* Vol. 89. No. 15 (2002) 158702.

10   Cabrera J L and Milton J G: "Delays, scaling and the acquisition of motor skill". Ms 2002.

11   Carpenter M G et al.: "Sampling duration effects on centre of pressure summary measures". *Gait and Posture* 13 (2001) 35-40.

12   Chechkin A et al.: "Stationary states of non-linear oscillators driven by Lévy-noise". *Chemical Physics* 284 (2002) 233 – 251.

13   Chiari L, Cappello A, Lenzi D  and Croce U Della: "An improved technique for extraction of stochastic parameters from stabilograms". *Gait and Posture* 12 (2000) 225 – 234.

14   Chiari L, Rocchi L and Cappello A: "Stabilometric parameters are affected by anthropometry and foot placement". *Clinical Biomechanics* 17 (2002) 666 – 677.

15	Chow C C and Collins J J: "Pinned polymer model of posture control". *Phys. Rev. E* 52 No. 1 (1995) 907 – 912.

16	Collins J J and De Luca C J: "Open-loop and closed-loop control of posture – a random walk analysis of center-of-pressure trajectories". *Exp. Brain. Res.* 95 (1993) 308 – 318.

17	Collins J J and De Luca C J: "Random walking during quiet standing". *Phys. Rev. Lett.* 73 No. 3 (1994) 764 – 767.

18	Delingières D *et al.*: "A methodological note on nonlinear time series analysis: Is the open- and closed-loop model of Collins and De Luca (1993) a statistical artifact?". *J. of Motor Behavior* 35 No. 1 (2003) 86 – 96.

19	Ding M and Yang W: "Distribution of the first return time in fractional Brownian motion and its application to the study of on-off intermittency". *Phys. Rev. E* 52. No. 1 (1995) 207 – 213.

20	Dingwell J B *et al.*: "Manipulating objects with internal degrees of freedom: Evidence for model-based control". *J. Neurophysiol.* 88 (2002) 222 – 235.

21	Engquist B and Schmid W (eds.): *Mathematics Unlimited – 2001 and Beyond*. Springer 2001.

22	Eurich C W and Milton J G: "Noise-induced transitions in human postural sway". *Phys. Rev. E* Vol. 54, No. 6 (1996) 6681 – 6684.

23	Flandrin P: "On the spectrum of fractional Brownian motions". *IEEE Transactions on Information Theory* 35 No. 1 (1989) 197 – 199.

24	Gagey P-M and Weber B: *Posturologie. Regulation et Dérèglements de la Station Debout*. 2e édition. Masson 1999. [Se även http://perso.club-internet.fr/pmgagey/ .]

25	Gravelle D C *et al.*:"Noise-enhanced balance control in older adults". *Neuroreport* 13 No. 15 (2002) 1853 – 1856.

26	Haken H: *Advanced Synergetics*. Springer 1983.

27	Haykin S: *Neural Networks. A Comprehensive Foundation*. 2. ed. Prentice Hall 1999.

28	Heagy J F, Platt N and Hammel S M: "Characterization of on-off intermittency". *Phys. Rev. E* 49 Nn. 2 (1994) 1140 – 1150.

29	Hsiao-Wecksler E T *et al.*: "Predicting the dynamic postural control response from quiet-stance behavior in elderly adults". *J. of Biomechanics* 36 (2003) 1327 – 1333.

30	Hänsler E: *Statistische Signale*. 2. Aufl. Springer 1997.

31	Ikeda K and Matsumoto K: "High-dimensional chaotic behavior in systems with time-delayed feedback". *Physica* D 29 (1987) 223- 235.

32	Jensen J J: *Self-Organized Criticality. Emergent Complex Behavior in Physical and Biological Systems*. Cambridge University Press 1998.

33	Jespersen D, Metzler R and Fogedby H C: "Lévy flights in external force fields: Langevin and fractional Fokker-Planck equations and their solutions." *Phys. Rev. E* 59 No. 2 (1999) 2736 – 2745.

34	Johnson R C: "Unicycles and bifurcations". *American J. Physics* 66 No. 7 (2002) 589 – 592.

35	Jones C K R T: "Whither applied nonlinear dynamics?" In Engquist B and Schmid W (2001).

36	José J V and Saletan E J: *Classical Dynamics. A Contemporary Approach*. Cambridge University Press 1998.

37	Kantz H, Kurths H and Mayer-Kress G (Eds.): *Nonlinear Analysis of Physiological Data*. Springer 1998.

38	Kantz H and Schreiber T: *Nonlinear Time Series Analysis*. Cambridge University Press 2000.

39	Kapteyn T S *et al.*: "Standardization in platform stabilometry being part of posturography". *Agressologie* 24 No.7 (1983) 321 – 326.

40   Khoo M C K: *Physiological Control Systems: Analysis, Simulation, and Estimation*. IEEE Press 2000.
41   Kitajo K *et al.*: "Behavioral stochastic resonance within the human brain". *Phys. Rev. Lett.* 90 No. 21 (2003) 218103.
42   Landau L D and Lifshitz E M: *Mechanics*. Pergamon Press 1976.
43   Lauk M *et al.*: "Human balance out of equilibrium: Nonequilibrium statistical mechanics in posture control". *Phys. Rev. Lett.* 80 No. 2 (1998) 413 – 416.
44   Liebovitch L S and Yang W: "Transition from persistent to antipersistent correlation in biological systems". *Phys. Rev. E* 56 No. 4 (1997) 4557 – 4566.
45   Lindner B *et al.*: "Effects of noise in excitable systems". *Physics Reports* 392 (2004) 321-424.
46   Linkenkaer-Hansen K: *Self-organized Criticality and Stochastic Resonance in the Human Brain*. [Diss.] Helsinki University of Technology 2002. ( http://lib.hut.fi/Diss/2002/isbn9512262177/ )
47   McIlroy W E and Maki B E: "Preferred placement of the feet during quiet stance: development of standardized foot placement for balance testing". *Clinical Biomechanics* 12 No. 1 (1997) 66-70.
48   Mandelbrot B B and Van Ness J W: "Fractional Brownian Motion, Fractional Noises and Applications". *SIAM Review* 10 (4) (1968) 422-437.
49   Mehta B and Schaal S: "Forward models in visuomotor control". *J. Neurophysiol.* 88 (2002) 942 – 953.
50   Metzler R and Klafter J: "The random walk's guide to anomalous diffusion: A fractional dynamics approach". *Physics Reports* 339, 1 (2000) 1- 77.
51   Miall R C *et al.*: "Is the Cerebellum a Smith predictor?" *J. of Motor Behavior* 25 No. 3 (1993) 203 – 216.
52   Milton J G: "Epilepsy: multistability in a dynamic disease". In Waellczek (2000) p. 374 – 386.
53   Milton J G *et al.*: "Controlling neurologcal disease at the edge of instability". Ms 2003.
54   Moreau L and Sontag E: "Balancing at the border of instability". *Phys. Rev. E* 68 (2003) 020901.
55   Moss F: "Stochastic resonance: looking forward". In Walleczek (2000) p. 236 – 256.
56   Nigg B M and Herzog W (eds.): *Biomechanics of the Musculo-skeletal System*. Wiley 1995.
57   Peterka R J: "Postural control model interpretation of stabilogram diffusion analysis". *Biological Cybernetics* 82 (2000) 335 – 343.
58   Press W H *et al.*: *Numerical Recipes in C++. The Art of Scientific Computing*. 2. ed. Cambridge University Press 2002.
59   Priplata A *et al.*: "Noise-enhanced human balance control". *Phys. Rev. Lett.* 89 No. 23 (2002) 238101.
60   Rangayyan R M: *Biomedical Signal Analysis. A Case-Study Approach*. IEEE Press 2002.
61   Rapp P E, Schmach T I and Mess A I: "Models of knowing and the investigation of dynamical systems". *Physica D* 132 (1999) 133 – 149.
62   Rosenblum M *et al.*: "Human postural control: Force plate experiments and modelling". In Kantz *et al.* (1998).
63   Schmid M *et al.*: "The sensitivity of posturographic parameters to acquisition settings". *Medical Engineering & Physics* 24 (2002) 623 – 631.
64   Smith O J M: "Closer control of loops with dead time". *Chemical Engineering Progress* 53 (1957) 217 – 219.
65   Soma R *et al.*: "1/*f* noise outperforms white noise in sensitizing baroreflex function in human brain". *Phys. Rev. Lett.* 97 No. 7 (2003) 078101.
66   Stépán G and Kollár L: "Balancing with reflex delay". *Mathematical and Computer Modelling* 31 (2000) 199 – 205.

67    Tettamanzi A and Tomassini M: *Soft Computing. Integrating Evolutionary, Neural, and Fuzzy Systems*. Springer 2001.
68    Thurner S, Mittermaier C and Ehrenberger K: "Change of complexity pattern in human posture during aging". *Audiology & Neuro Otology* 7 (2002) 240 – 248.
69    Saichev A I and Woyczyñski W A: *Distributions in the Physical and Engineering Science* (Vol. 1). Birkhäuser 1997.
70    Walleczek J (ed.): *Self-organized Biological Dynamics and Nonlinear Control*. Cambridge University Press 2000.
71    Winter D A *et al.*: "Ankle muscle stiffness in the control of balance during quiet standing". *J. Neurophysiol.* 85 (2001) 2630 – 2633           .
72    Winter D A *et al.*: "Motor mechanism of balance during quiet standing". *Journal of Electromyography and Kinesiology* 13 (2003) 49 – 56.
73    Winter D A *et al.*: "Stiffness control of balance in quiet standing". *J. Neurophysiol.* 80 (1998) 1211 – 1221.
74    Wolpert D M, Ghahramani Z  and Jordan M I : "An internal model for sensimotor integration". *Science* 269 (1995) 1880 – 1882.
75    Wolpert D M: "Computational approaches to motor control". *Trends in Cognitive Sciences* 1 No. 6 (1997) 209 – 216.
76    Zatsiorsky V M and Duarte M: "Rambling and trembling in quiet standing". *Motor Control* 4 (2000) 185 – 200.

*Illustrations*

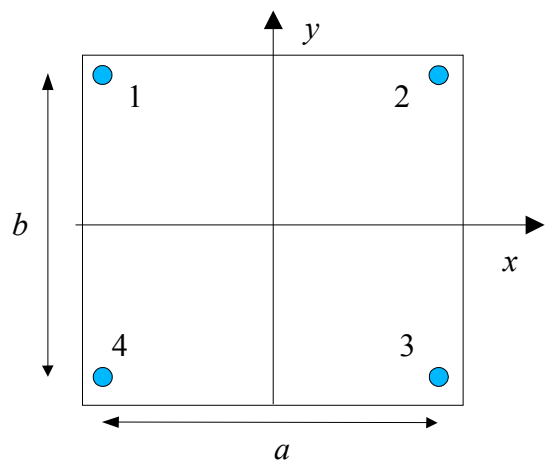

Fig. 1: Force place (schematic) seen from above, placements 1 – 4 of the force transducers indicated.

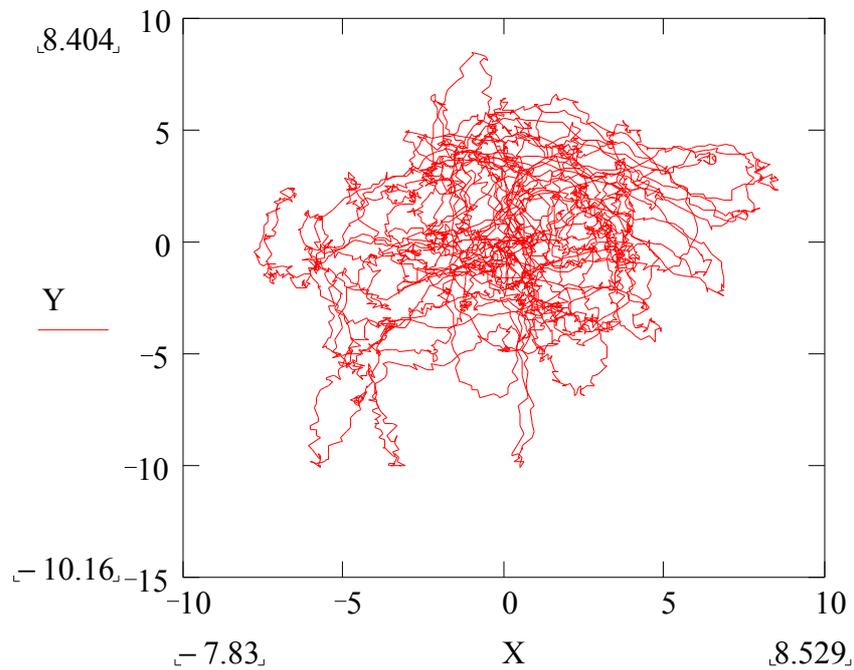

Fig. 2: Posturogram showing the trace of the center-of-pressure (COP). Length unit = mm.

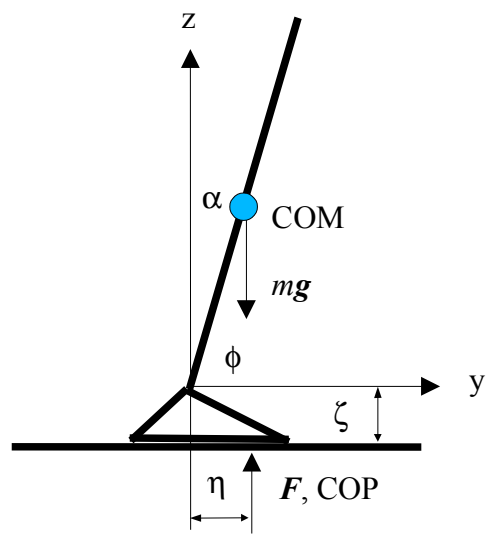

Fig. 3: The human inverted pendulum model. This model is a good approximation when the standing person applies the so called "ankle strategy" which typically is the case.

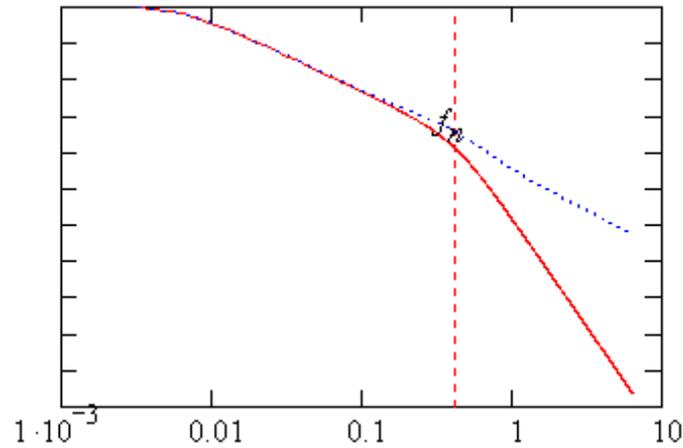

Fig. 4: Theoretical (smoothed) spectra for COP-A/P coordinate (dotted line) and the COM-A/P coordinate for the model presented by equ (13).

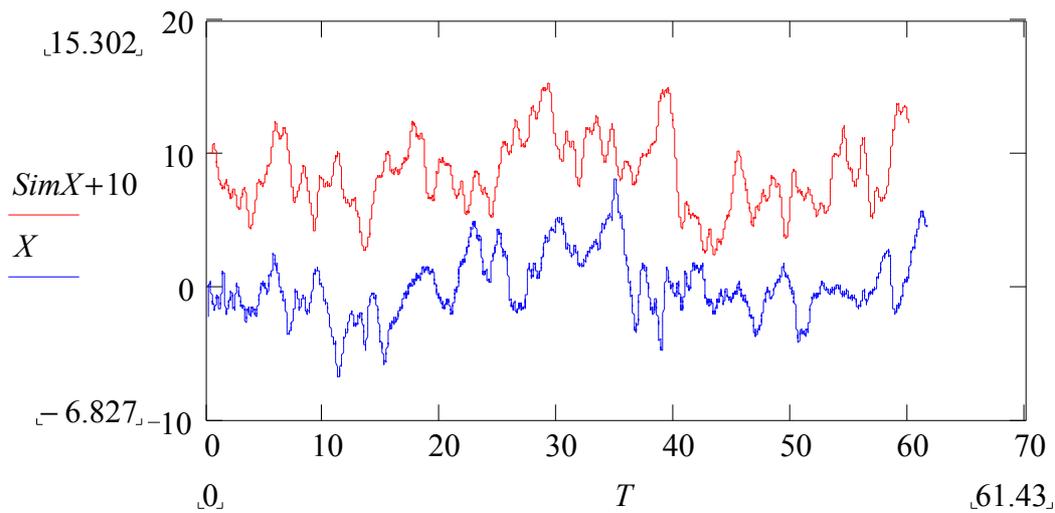

Fig. 5: ARMA-simulated COP-M/L trace (upper curve) according to equ (16) and experimental curve (below). Length unit = mm, time unit = seconds.